\input harvmac.tex      

%
%
%

\def\tilde{\widetilde}
\def\bar{\overline}
\def\hat{\widehat}
\def\*{\star}
\def\[{\left[}
\def\]{\right]}
\def\({\left(}		
\def\){\right)}

%
%
\def\zb{{\bar{z} }}
\def\frac#1#2{{#1 \over #2}}
\def\inv#1{{1 \over #1}}

\def\d{\partial}

\def\2pi{\hbox{$2\pi i$}}

\def\dsl{\raise.15ex\hbox{/}\kern-.57em\partial}
\def\Dsl{\,\raise.15ex\hbox{/}\mkern-.13.5mu D}
%
%

\def\al{\alpha}
\def\ep{\epsilon}

%
%
	\def\CB{{\cal B}}

		\def\CL{{\cal L}}
		\def\CO{{\cal O}}

\def\2pi{\hbox{$2\pi i$}}

\def\dsl{\raise.15ex\hbox{/}\kern-.57em\partial}
\def\Dsl{\,\raise.15ex\hbox{/}\mkern-.13.5mu D}
%
%
%
\font\numbers=cmss12
\font\upright=cmu10 scaled\magstep1
\def\stroke{\vrule height8pt width0.4pt depth-0.1pt}
\def\topfleck{\vrule height8pt width0.5pt depth-5.9pt}
\def\botfleck{\vrule height2pt width0.5pt depth0.1pt}
\def\Zmath{\vcenter{\hbox{\numbers\rlap{\rlap{Z}\kern
0.8pt\topfleck}\kern
2.2pt
                   \rlap Z\kern 6pt\botfleck\kern 1pt}}}
\def\Qmath{\vcenter{\hbox{\upright\rlap{\rlap{Q}\kern
                   3.8pt\stroke}\phantom{Q}}}}
\def\Nmath{\vcenter{\hbox{\upright\rlap{I}\kern 1.7pt N}}}
\def\Cmath{\vcenter{\hbox{\upright\rlap{\rlap{C}\kern
                   3.8pt\stroke}\phantom{C}}}}
\def\Rmath{\vcenter{\hbox{\upright\rlap{I}\kern 1.7pt R}}}
\def\Z{\ifmmode\Zmath\else$\Zmath$\fi}
\def\Q{\ifmmode\Qmath\else$\Qmath$\fi}
\def\N{\ifmmode\Nmath\else$\Nmath$\fi}
\def\C{\ifmmode\Cmath\else$\Cmath$\fi}
\def\R{\ifmmode\Rmath\else$\Rmath$\fi}


\def\del{\partial}

\def\bs{{\hat{\beta}^2}}

\Title{CLNS 1451, NSF-ITP-97-022, hep-th/9703085}
{\vbox{\centerline{Purely Transmitting Defect Field Theories }
 }}

\bigskip
\bigskip

\centerline{Robert Konik}
\medskip\centerline{Newman Laboratory}
\centerline{Cornell University}
\centerline{Ithaca, NY  14853}
\bigskip\bigskip
\centerline{Andr\'e LeClair\foot{On leave from Cornell University}}
\medskip\centerline{Institute for Theoretical Physics}
\centerline{University of California}
\centerline{Santa Barbara, CA 93106-4030}

\bigskip\bigskip

\vskip .3in

We define an infinite class of integrable theories with a defect
which are formulated as chiral defect perturbations of a
conformal field theory.  Such theories can be 
interacting in the bulk, and  are purely transmitting
through the defect.  The examples of the sine-Gordon theory and Ising
model are worked out in some detail.

\Date{3/97}

%
%
%
%
%
%
%
%
%

\newsec{Introduction} 

The class of integrable quantum field theories has grown 
dramatically in the last several years due primarily to the idea 
of Zamolodchikov that defines such theories as perturbations
of known conformal field theories \ref\rcpt{A. Zamolodchikov,
Int. Journ. Mod. Phys. A4 (1989) 4235.}.   This approach has been
generalized to theories on the half-line with non-trivial
boundary interactions \ref\rgosh{S. Ghoshal and A. Zamolodchikov,
Int. J. Mod. Phys. A 9 (1994) 3841.}.  

Another class of theories with many interesting applications
has a single impurity or defect not at the boundary but
embedded in the sample.  
Such a theory can be described by the Euclidean action,
\eqn\eIi{
S = S_{\rm bulk} + g \int dt ~ \CO (0,t) , }
where the bulk action is the integral of the Lagrangian density,
$S_{\rm bulk} = \int dx dt \CL$, and the spatial variable 
$x$ varies over the whole real line.  The perturbation by the
defect operator $\CO$ at $x=0$ modifies the Hamiltonian as follows:
\eqn\eIii{
H = H_{\rm bulk} + g ~  \CO (0) .}
For a general defect operator $\CO$, translation invariance is
broken.  Thus particles interacting with the defect can be {\it both}
transmitted or reflected.  
The integrability of such 
defect field theories was studied in \ref\rmuss{G. Delfino, 
G. Mussardo, P. Simonetti, Nucl. Phys. B432 (1994) 518.}.
Unfortunately, non-trivial solutions to the Yang-Baxter-like
constraints involving both reflection and transmission are
severely limited: for diagonal bulk scattering, it was shown
in \rmuss\ that non-trivial reflection and transmission
are only allowed in theories that are free in the bulk\foot{We
verified that the same occurs in a non-diagonal bulk scattering
theory, specifically the sine-Gordon theory (though with no
defect degrees of freedom).}.
Thus integrable impurity problems that are interacting in
the bulk are of only two types:  purely reflecting, which
is the same as a boundary theory, and purely transmitting. 

In this paper, using conformal perturbation theory, 
 we argue that the purely transmitting theories
arise from actions of the kind \eIi\ where the defect operator
$\CO$ is chiral. 
We describe how to associate an integrable chiral defect theory to
every known integrable bulk perturbation of a conformal field theory. 
We show that if the bulk theory is a massless conformal field theory
then the chiral defect theory can be mapped onto a massless boundary
field theory.  Interestingly, in general this map requires one to
introduce defect degrees of freedom.  
Genuinely new models occur when the bulk is massive,
since in this case the theory cannot be folded onto a known boundary
theory.   These general features are illustrated with the examples
of the chiral defect sine-Gordon theory and a chiral defect
perturbation of the Ising model.

\newsec{Integrals of Motion and Conformal Perturbation Theory}

We study first the general features of integrals of motion in
theories with a defect defined by an action \eIi.  
The defect separates space into two regions, $x<0$ and $x>0$, 
and one must distinguish the fields in these regions.  For any
field $\Psi (x,t)$, 
\eqn\eIiii{\eqalign{
\Psi (x,t) &= \theta(x) \Psi^{(+)} (x,t) + \theta (-x) \Psi^{(-)} (x,t) \cr
\Psi (0,t) &\equiv \inv{2} 
\( \Psi^{(+)} (0,t) + \Psi^{(-)} (0,t) \). 
\cr }}

Suppose we are given a bulk conserved current $J_\mu$, satisfying
$\d_t J_t - \d_x J_x =0$ where $x \neq 0$.  This current will lead
to a conserved charge in the presence of the defect if the following
boundary condition is satisfied
\eqn\eIiv{
J_x^{(+)} (0,t) - J_x^{(-)} (0,t) = \d_t \Theta (t) , }
for some operator $\Theta$.  A conserved charge satisfying 
$\d_t Q = 0$ is then constructed as follows: 
\eqn\eIv{
Q = \Theta (t) + \int_{-\infty}^\infty dx  ~ J_t . }

\def\wb{{\bar{w}}}

We now take the bulk theory to be defined by a bulk perturbation 
of a conformal field theory (CFT):
\eqn\eIvi{
S_{\rm bulk} = S_{CFT} + \lambda \int dx dt ~ \Phi (x,t) . }
The CFT must be a specific defect CFT, i.e. the boundary
conditions at the defect must preserve the conformal invariance.
See section 3.  
For a given CFT, certain perturbing fields $\Phi$ lead to 
an integrable quantum field theory \rcpt. 
We henceforth assume that based on the specific CFT of the bulk,
$\Phi$ has been chosen to define an integrable theory. 
This implies there are an infinite number of conserved currents. 
In conformal perturbation theory, these currents have the following
description.  When $\lambda=0$, the perturbing field factorizes
into left and right moving parts: $\Phi(z, \zb)  = \phi_L (z) 
\phi_R (\zb) $, where $z = t+ ix$, $\zb = t-ix$.  If the theory
is integrable, there exists an infinite number of chiral fields
(for $\lambda = 0$), $\{ J_L (z), J_R (\zb) \}$ with the following
operator product expansion with the perturbing operator:
\eqn\eIvii{\eqalign{
\phi_L (w) J_L (z) &= \ldots \inv{w-z} \d_z H_L (z) + \ldots 
\cr
\phi_R (\wb ) J_R (\zb ) &= \ldots \inv{\wb - \zb } \d_\zb H_R (\zb ) + \ldots 
, 
\cr }}
i.e. the residue of the simple pole is a total derivative.  
This ensures the conservation of the following charges:
\eqn\eIviii{
\eqalign{
Q &= \int dz J_L + \lambda \int d\zb \( H_L \phi_R \) \cr
\bar{Q}  &= \int d\zb J_R + \lambda \int dz \( H_R \phi_L \) . 
\cr}}

With the right choice of defect perturbation, the operator product
expansion \eIvii\ can ensure integrals of motion in the defect
theory.  Some analysis  of the integrals of motion
shows that in general the choice of the local defect operator
$\CO = \Phi$ does not lead to an integrable theory\foot{In some cases
a change of basis and/or an intricate folding of the theory 
can relate it to an integrable
boundary theory; see section 5.}.  
However a chiral perturbation does.  Namely, we consider 
the action
\eqn\eIix{
S = S_{\rm bulk} + g \int dt ~ \phi_L (0,t) , }
where $S_{\rm bulk}$ is as in \eIvi, where $\Phi = \phi_L \phi_R$.
(We can just as well perturb with $\phi_R $ rather than $\phi_L$.) 

There 
are subtle issues concerning the locality of a chiral perturbation since
for a general CFT the integrable chiral fields have fractional 
Lorentz spin.  We will address this issue for the sine-Gordon
theory below.   Alternatively, this problem 
 can perhaps be cured by multiplying 
$\phi_L $ by a discrete degree of freedom which can be 
interpreted as a zero mode of a right-moving field.  These
discrete degrees of freedom may be the origin of the defect
degrees of freedom introduced below.   For instance,
see the treatment of the free fermion in section 6.

First set $\lambda = 0$. 
To first order in perturbation theory
one has, 
\eqn\eIx{
J^{(+)}_L (0,t) - J^{(-)}_L (0,t) = \lim_{\ep \to 0^+} 
g \int dt' \phi_L (0,t') \( J_L (\ep, t) - J_L (-\ep, t) \). }
Using the operator product expansions \eIvii, 
\eqn\eIxi{\eqalign{
J^{(+)}_L (0,t) - J^{(-)}_L (0,t) &= 4\pi i g\d_t H_L \cr
J^{(+)}_R (0,t) - J^{(-)}_R (0,t) &= 0. \cr }} 
This implies the existence of {\it two} conserved charges 
for every $J_{L,R}$: 
\eqn\eIxii{\eqalign{
Q_L &= 4\pi g H_L + \int dx J_L ;\cr
Q_R &= \int dx J_R . \cr}}

In specific models, one can usually apply scaling arguments to rule out
higher order corrections in $g$ that could spoil the
conservation laws.  In the massless case, the folding construction
of the next section indicates that if the boundary version of the
conserved charges are exact to first order in $g$, then so are the 
above charges in the defect theory. 

For the case of the energy momentum tensor, the above analysis
indicates that both left and right moving momentum and thus both
the usual energy and momentum are conserved.  This means that there can be
no reflection of particles off the boundary, as this would violate
momentum conservation.  Therefore, integrable theories defined by
\eIix\ are purely transmitting, at least in the massless case,
with generally non-trivial 
transmission S-matrices through the defect.  

Now we consider the theory 
\eIix\ with $\lambda \neq 0$.  
One might expect that since the bulk perturbation couples left 
with right movers, these theories will have both reflection
and transmission, and if the bulk S-matrix is non-trivial,
the remarks made in the introduction would suggest this theory
cannot be integrable.  
To settle the issue, 
one must check that the additional bulk
terms in \eIviii\ do not spoil the boundary condition \eIiv. 
We will address these points in the sine-Gordon example below. 
There we will argue, though not
conclusively,  that the theory continues to be integrable
for $\lambda \neq 0$, and will propose transmission S-matrices for
the model.

\newsec{Folding for a massless bulk theory}

\subsec{Defect Conformal Field Theory}

In conformal field theory,  boundary conditions at the defect
which preserve the conformal symmetry satisfy
\eqn\econf{\eqalign{
T^{(+)} (0,t) - T^{(-)} (0,t) &= 0 \cr 
\bar{T}^{(+)} (0,t) - \bar{T} ^{(-)} (0,t) &= 0,  \cr 
}}
where $T(z) , \bar{T} (\zb )$ are the chiral and anti-chiral
components of the energy-momentum tensor.  Using the analyticity
one can in general fold this theory onto the half-line
$x>0$ as follows.  Define boundary fields for $x>0$ as follows:
\eqn\eefold{\eqalign{
T_B (x,t) &=  T^{(+)} (x,t) ,  ~~~~~\bar{T}_B (x,t) = T^{(-)} (-x,t), \cr 
\bar{T}'_B (x,t) &=  \bar{T}^{(+)} (x,t) ,  
~~~~~{T}'_B (x,t) = \bar{T}^{(-)} (-x,t) . \cr}}
The fields $T_B , T'_B$ are functions of $t+ix$ whereas
$\bar{T}_B , \bar{T}'_B $ are functions of $t-ix$ for $x>0$. 
The boundary conditions \econf\ imply
\eqn\econfii{\eqalign{
T_B (0,t) - \bar{T}_B (0,t) &= 0 \cr
T'_B (0,t) - \bar{T}'_B  (0,t) &= 0 ,\cr
}}
thus in the region $x>0$ one has {\it two} boundary conformal field
theories which are decoupled, corresponding to $T_B$ verses $T'_B$. 
Boundary conformal field theories were studied in \ref\rcardy{J. L. Cardy,
Nucl. Phys. B275 (1986) 200-218.}. 
We conclude that defect conformal field theories are characterized
by the tensor product of two conformal boundary conditions. 
This fact was used for instance in \ref\raff{M. Oshikawa and I. 
Affleck, Phys. Rev. Lett. 77 (1996) 2604.}.

\subsec{Chiral Perturbations}

When the bulk theory is massless, i.e. $\lambda =0$ in \eIvi, 
then the defect theory \eIix\ can be reformulated as a 
boundary theory on the half-line by the general folding
procedure above.  This has already been used in a number
of contexts, e.g. \ref\rfold{E. Wong   
and I. Affleck, Nucl. Phys B417 (1994) 403.} \ref\rfold'{P. Fendley,
A.W.W. Ludwig, H. Saleur, Phys. Rev. Lett. (1995) 3005.}. 
Here the situation is even simpler, since in the 
perturbed chiral 
defect theory, only left-moving fields are coupled to the
defect. Thus, the boundary theory defined by 
$T'_B , \bar{T}'_B$ decouples from the defect.
Let $\varphi^{(\pm)}_L$ denote the components of an arbitrary
chiral field  
$\varphi_L$ on either side of the defect, as in \eIiii. 
Now, let us define a boundary theory in the region $x\geq 0$, where
for each field  $\varphi_L$ of the defect theory we associate the 
boundary theory fields $\varphi_{L,R}$: 
\eqn\eeIIi{
\varphi_L (x,t) = \varphi_L^{(+)} (x,t) , 
~~~~~\varphi_R (x,t) = \varphi^{(-)}_L (-x,t),
~~~~~x>0.}
Since the bulk is massless, $\varphi_L (x,t) = \varphi_L (z)$, 
$\varphi_R (x,t) = \varphi_R (\zb )$.  The boundary condition 
\eIxi\ now becomes in the boundary theory:
\eqn\eeIIii{
J_L (0,t) - J_R (0,t) = 4\pi i g ~  \d_t H_L .}
The above condition assures the existence of a conserved charge
in the boundary field theory \rgosh:
\eqn\eeIIiii{
Q = \( \int_0^\infty dx ( J_L + J_R) \)  +  4 \pi g H_L . }
$Q$ is the direct map of the charge $Q_L$ of the defect theory
to the boundary theory, i.e. 
\eqn\eeIIiv{
Q_L = \int_{-\infty}^0 dx J^{(-)}_L + \int_0^\infty dx J^{(+)}_L 
+ 4\pi g H_L = Q . }

\def\t{\theta}

\newsec{Defect S-matrices}

Above we showed that interacting integrable theories with a defect
are purely transmitting through the defect.  We now derive
the algebraic equations satisfied by the transmission S-matrices. 
The massive and massless cases should be treated differently. 

\subsec{Massive case} 

\def\ap{A^{(+)}}
\def\am{A^{(-)}}
\def\tt{{\tilde{T}}}

Consider the bulk theory to be an integrable theory as in 
\eIvi\ with $\lambda \neq 0$, so that the spectrum consists of
massive particles.  Let $A_a (\t )$ denote the formal 
Faddeev-Zamolodchikov operators for particles of type `a'
and rapidity $\t$, where as usual the momentum is $m\sinh \t$. 
We distinguish particles to the left (right) of the defect as 
$A^{(-)}_a $ ($ A^{(+)}_a$). 
They are defined to satisfy the exchange relation 
\eqn\eIIi{
A^{(\pm)}_a (\t_1 ) A^{(\pm)}_b (\t_2 ) 
= S_{ab}^{dc} (\t_1 - \t_2 ) A^{(\pm)}_c (\t_2 ) A^{(\pm)}_d (\t_1 ) , }
where $S_{ab}^{dc}$ is the bulk S-matrix satisfying 
Yang-Baxter, crossing and unitarity equations. 
Let us as before place the defect at $x=0$ and then introduce a defect 
operator $D_\al$, with defect degrees of freedom $\al$ taking values in
some set.  The scattering of particles through the defect is then
described by the following exchange relations:
\eqn\eIIii{\eqalign{
D_\al ~ \ap_a (\t) &= T_{\al a}^{\beta b} (\t ) \am_b (\t ) D_\beta \cr
\am_a (\t ) D_\al  &= \tt_{a\al}^{b\beta} (-\t ) D_\beta \ap_b (\t ) . 
\cr }}
The transmission scattering matrices $T$ and $\tt$ are not necessarily
identical because parity symmetry is broken. 

Consistency of \eIIii\ implies the unitarity condition
\eqn\eIIiii{
T_{\al a}^{\beta b} (\t ) \tt_{b\beta}^{c\gamma} (-\t ) = \delta_a^c 
\delta_\alpha^\gamma . }
We will also require the crossing symmetry relation
\eqn\eIIiv{
\tt_{a\al}^{b\beta} (\t ) = T_{\al \bar{b}}^{\beta \bar{a}} (i\pi - \t ), }
where $\bar{a}$ is the charge conjugate of $a$. 

Associativity of the algebra \eIIii\ demands $T$ and $\tt$ satisfy
the defect Yang-Baxter equations (DYB): 
\eqna\eIIv
$$\eqalignno{
S_{ab}^{dc} (\t_{12} ) T_{\al c}^{\beta e} (\t_2 ) T_{\beta d}^{\gamma f}
(\t_1 ) 
&= T_{\al a}^{\beta c} (\t_1 ) T_{\beta b}^{\gamma d} (\t_2 ) 
S_{cd}^{fe} (\t_{12}) &\eIIv {a} \cr
S_{ab}^{dc} (\t_{12} ) \tt_{d\al}^{e\beta} (-\t_1 ) \tt_{c\beta}^{f\gamma}
(-\t_2 ) 
&= \tt_{b\al}^{c\beta} (-\t_2 ) \tt_{a\beta}^{d\gamma} (-\t_1 ) 
S_{dc}^{ef} (\t_{12}) 
&\eIIv {b} , \cr }$$
where $\t_{12} = \t_1 - \t_2 $.  
If $S_{ab}^{dc} = S_{ba}^{cd}$, then upon relabeling $\t_1 \to -\t_2 $,
$\t_2 \to - \t_1 $, one sees that $\tt_{a\al}^{b\beta}$ 
satisfies the same DYB equation as $T_{\al a}^{\beta b}$.  

A final constraint arises from hermiticity of the Hamiltonian. 
The transmission S-matrices are understood as the matrix elements
\eqn\eIIvc{\eqalign{
T_{\al a}^{\beta b} (\t ) &= 
{}_{out}^L \langle \t , b, \beta | \t , a, \al \rangle_{in}^R \cr
\tt_{ a \al }^{ b \beta } (- \t ) &= 
{}_{out}^R \langle \t , b, \beta | \t , a, \al \rangle_{in}^L \cr
}}
where here $L,R$ denotes to the left or right of the defect.  
Hermiticity of the Hamiltonian demands 
\eqn\eIIvd{
\tt_{a\al}^{b\beta} (\t ) = \( T_{\beta b}^{\al a} (-\t ) \)^*  . }

\subsec{Massless case} 

Now we take the bulk theory to be massless, i.e. $\lambda = 0$ in 
\eIvi.  Only left-moving particles couple to the defect, so we only
deal with these left-movers and denote them $A_a^L (k)$, for
$k<0$.  It is convenient to parameterize the momentum with a 
rapidity $\t$, $k_L = -\frac{\mu}{2} e^{-\t}  $, where $\mu$ is some scale. 
The scattering with the defect is described by
\eqn\eIIvi{
D_\al A_a^L (\t ) = T_{\al a}^{\beta b} (\t ) A_b^L (\t ) D_\beta , }
and the DYB equation is the same as \eIIv{a}.  The massless bulk
scattering is described by 
\eqn\eIIvii{
A^{L}_a (\t_1 ) A^{L}_b (\t_2 ) 
= S_{ab}^{dc} (\t_1 - \t_2 ) A^{L}_c (\t_2 ) A^{L}_d (\t_1 ) , }

In the last section, for the massless case we showed at the level of
fields how the problem can be mapped onto a boundary field theory. 
We now describe how this folding is performed in the scattering theory. 
Consider a massless boundary field theory on the half-line $x\geq 0$. 
This theory has both left and right moving particles.  Parameterize 
the left-moving momenta as $k_L = - \frac{\mu}{2}  e^{-\t_L} $, and the 
right-moving momenta as $k_R = \frac{\mu}{2} e^{\t_R} $, and introduce
operators $A_a^L (\t_L) $ and $A_a^R (\t_R )$.  
(We will not display the subscripts $L, R$ on $\t_{L,R}$ when the
meaning is clear.) 
The scattering in the bulk is as in \eIIvii, and the same equation
also holds for $A^L \to A^R$.  By scale invariance, the $L-R$ scattering
is independent of momentum.  However non-trivial constant scattering
is allowed, and will be important in the sequel.  Thus we suppose
\eqn\eIIviii{
A_a^R (\t_1 ) A_b^L (\t_2 ) = B_{ab}^{dc} A_c^L (\t_2) A_d^R (\t_1 ) , }
where $B_{ab}^{cd}$ is a matrix of constants satisfying a braiding relation.
The interaction with the boundary is encoded in a reflection
matrix.  Assuming the boundary to have no degrees of freedom, we
introduce a boundary operator $\CB$ satisfying:
\eqn\eIIix{
\CB A_a^L (\t ) = R_a^b (\t ) \CB A_b^R (-\t ) . }
Consistency of \eIIvii\ and \eIIix\ demands the massless boundary 
Yang-Baxter (BYB) equation: 
\eqn\eIIx{
R_a^c (\t_1 ) B_{cb}^{c'b'} R_{b'}^d (\t_2 ) S_{dc'}^{ef} (\t_{12})
= S_{ab}^{dc} (\t_{12} ) R_c^{e'} (\t_2 ) B_{e'd}^{ed'} R_{d'}^f (\t_1 ) . }

When the L-R scattering is trivial, i.e. if 
$B_{ab}^{a'b'} = \delta_a^{a'} \delta_b^{b'} $, then the massless
BYB equation \eIIx\ is identical to the DYB equation
with no defect degrees of freedom.  Namely, with no defect degrees of
freedom, $T_{\al a}^{\beta b} = T_a^b$, and a solution to \eIIx\ is
a solution to \eIIv{a} with $T_a^b (\t ) = R_a^b (\t )$.  
Thus, a massless defect theory can be mapped onto a massless boundary
theory by identifying the transmission and reflection S-matrices
only when L-R scattering is trivial.  

In general, one does not have the freedom to set the L-R scattering to
$1$.  Given a massive theory in the bulk, with $k=m\sinh \t$, the massless
limit $m\to 0$ is defined by letting $\t = \t_L -\al $, for left movers,
and $\t = \t_R + \al$ for right-movers, and in both cases letting 
$\al \to \infty$ keeping $me^\al = \mu$ held fixed. 
If $S_{ab}^{dc} (\t )$ is the massive S-matrix, then in this limit
$L-L$ and $R-R$ scattering are both given by the same $S$, whereas
$L-R$ scattering is given by 
\eqn\eIIxii{
B_{ab}^{dc} = S_{ab}^{dc} (-\infty) . }
Given a non-trivial $B$, one can try to establish a map between
massless defect and boundary theories by introducing defect
degrees of freedom.  Indeed, in the example of the sine-Gordon
theory treated in the next section, we find that by properly
introducing defect degrees of freedom, we can find a one-to-one
correspondence between solutions of DYB and massless boundary
Yang-Baxter equations. 

\def\rh{\hat{R}} 

In light of this, we can propose a general approach to the 
massive case of purely transmitting defect theories.  In the
massive case, folding is not possible.  For instance, in the massive
case, the BYB equation reads \rgosh ,
\eqn\eIIxiii{
\rh_a^c (\t_2 ) S_{bc}^{b'c'} (\t_1 + \t_2 ) 
\rh_{b'}^d (\t_1 ) S_{c'd}^{fe} (\t_{12})
= S_{bd}^{ca} (\t_{12} ) \rh_c^{e'} (\t_1 ) S_{de'}^{d'e} (\t_1 + \t_2 )
\rh_{d'}^f (\t_2 ) , }
where $\rh_a^b$ is the massive reflection S-matrix.  
It is clear that only in the massless limit \eIIx\ is there a chance
of mapping BYB onto DYB.  
Our approach to the massive case will be as follows.  Take the
massless limit of the defect theory, and map it onto a known
solution of a massless boundary field theory.  If the $L-R$ 
scattering of latter is non-trivial, the map can only be established
by introducing defect degrees of freedom.  
Finally, assume the massive case has the same defect 
degrees of freedom, and solve the defect Yang-Baxter equation
for the massive case.  We will illustrate this in the next
section with the sine-Gordon theory. 

\newsec{Defect sine-Gordon theory} 

In this section we treat the chiral defect perturbation
of the sine-Gordon theory, defined by 
\eqn\eIIIi{
S = \int_{-\infty}^\infty dx dt \( \inv{2} (\d_\mu \Phi)^2 
+ \lambda \cos \beta \Phi \)  +  g \int dt \cos \beta \phi_L (0,t) , }
where $\phi_L$ is the left-moving component of the scalar field,
$\Phi = \phi_L + \phi_R$.   
The bulk sine-Gordon theory has a well-known infinite set of
local conserved charges with spin equal to an odd integer.  
The general arguments of section 2 indicate that the above  
{\it chiral defect sine-Gordon theory}, at least in
the massless limit $\lambda = 0$,  is also integrable.

At the level of the action \eIIIi\ one can fold the massless theory
as described in section 3 and explicitly obtain the boundary
sine-Gordon theory (BSG).  From  
the components $\phi_L^{(\pm)}$ of the field $\phi_L$ in the
defect theory, define the boundary theory fields 
$\varphi_L (x,t) = \phi_L^{(+)} (x,t) $, $\varphi_R (x,t) 
= \phi_L^{(-)} (-x,t) $ for $x>0$, as in \eeIIi.  
In \eIIIi\ since $\phi_L (0,t) = ( \phi_L^{(+)} (0,t) + \phi_L^{(-)} (0,t) 
)/2$,  in the map to the boundary theory one finds
\eqn\map{
\int dt ~ \cos \beta \phi_L (0,t)  \to  \int dt ~ 
\cos \( \beta ( \varphi_L (0,t) 
+ \varphi_R (0,t) )/2 \)  =  \int dt ~ \cos ( \beta \Phi_B /2 ) , }
where now the boundary field 
$\Phi_B = \varphi_L + \varphi_R$ is the boundary sine-Gordon field. 
The complete action \eIIIi\ becomes the BSG theory, as defined in 
\rgosh.   
We thus argue that in the massless case, if the boundary sine-Gordon
conserved currents are exact to first order in $g$, then so are the
conserved currents in the defect theory.   

The problem with the apparent non-locality of the chiral perturbation in 
\eIIIi\ can be resolved in the following way.  Chiral conformal fields
satisfy the braiding relations:
\eqn\ebraid{
e^{i a \phi_L (x,t) } e^{i b \phi_L (y,t) } 
= e^{\pm i  ab/4} 
e^{i b \phi_L (y,t) } e^{i a \phi_L (x,t) } 
~~~~~ {\rm for} ~ \( {x>y \atop x<y} \). }
Writing $\phi_L (0,t) = \lim_{\ep \to 0} ( \phi_L(\ep , t) + \phi_L (-\ep ,t) 
)/2 $,  one finds that
any field $\exp (i a \phi_L (0,t) )$ at the defect
 is local with respect to 
the perturbation $\exp( \pm i \beta 
( \phi_L (\ep ,t ) + \phi_L (-\ep , t) )/2 )$, since from the braiding 
relation \ebraid, the phase from $\phi_L (\ep, t)$ is canceled by
the phase from $\phi_L (-\ep, t)$. 
 
Finally, it is interesting to consider taking not a chiral defect
perturbation, but a local one in \eIIIi.  Namely, let us
replace $\cos \beta \phi_L$ with $\cos (\beta \Phi /2)$ , where
$\Phi = \phi (z) + \bar{\phi} (\zb )$ is the local SG field. 
By making the change of basis $\phi_{\pm}(x,t) = \( \phi (x,t) \pm
\bar\phi (-x,t) \)/2$, the interaction is seen to depend only on 
$\exp (\beta \phi_+)$.  As $\phi_{\pm}$ are chiral, $\exp (\beta \Phi /2)$
is integrable if $\exp (\beta \phi )$ is integrable.
We can also see this via a folding into a boundary theory.
As described in section 3, when $g=0$ this theory can be folded
onto two decoupled boundary conformal field theories for $x>0$.
Namely, following 
\eefold, define boundary fields 
$\varphi_L (x,t) = \phi^{(+)} (x,t)$, 
$\varphi_R (x,t) = \phi^{(-)} (-x,t)$, 
$\varphi'_L (x,t) = \bar{\phi}^{(-)} (-x,t)$, 
$\varphi'_R (x,t) = \bar{\phi}^{(+)} (x,t)$. 
Unlike the purely chiral perturbation \eIIIi, now  
the defect interaction couples the two boundary conformal field
theories:   
\eqn\ecouple{
\cos (\beta \Phi (0,t)/2 ) \to \cos (\beta (\Phi_B  + \Phi'_B )/4 ), }
where
 now $\Phi_B = \varphi_L + \varphi_R$,
and $\Phi'_B = \varphi'_L + \varphi'_R$.  
Since the combination $\Phi_B - \Phi'_B$ decouples from the defect,
what remains is again a single boundary SG theory.   So again a local
defect perturbation is seen to be integrable, but now via an
intricate folding.  Of course, these arguments only hold for $\lambda = 0$.

\subsec{Scaling Analysis of Integrability}

We denote the set of charges for the sine-Gordon by
\eqn\eIVi{\eqalign{
Q_s = \int dx J_s - \lambda \int dx R_s, \cr
\bar{Q}_s = \int dx \bar{J}_s - \lambda \int dx \bar {R}_s\cr}}
where $J_s$ ($\bar{J}_s$) has spin s (-s), s even.   
The local fields 
$J_s, R_s, \bar{J}_s,$ and $\bar{R}_s$ satisfy
\eqn\eIVii{\eqalign{
\del_{\bar z} J_s = \lambda \del_z R_s , \cr
\del_z \bar{J}_s = \lambda \del_{\bar z}\bar{R}_s.\cr}}
In the presence of a defect line at $x=0$, the most general modification
the left moving fields, $J_s$ and $R_s$, can undergo is
\eqn\eIViii{\eqalign{
J_s^{(+)}  (0) - J_s^{(-)}  (0) = \sum^\infty_{n=1}  g^n J^{n}_s ,\cr
R_s^{(+)} (0) - R_s^{(-)} (0) 
= \sum^\infty_{n=1}  g^n R^{n}_s .\cr}}
(We will assume  that no terms dependent on $\lambda$  appear on the
RHS of \eIViii .) 

The possible non-zero terms on the RHS of \eIViii\ can be partially
determined by a scaling analysis.  The scaling dimension of 
$g$ is $d(g) = 1 - \bs/2$, $d(J_s) = s$, and 
$d(R_s) = s-2 +\bs$.  
(For convenience we have defined $\hat{\beta} = \beta/\sqrt{4\pi}$.)
Since $J_s^n $ arises to n-th order in perturbation theory, it must
come from the operator product expansion of $J_s$ with 
$\cos^n (\beta \phi_L )$.  Thus 
$d(J_s^n) = d(e^{i r \beta  \phi_L} \CO_s) $, where $r$ is an integer with  
$|r| \leq n $, and $\CO_s $ is a local field of integer dimension $l$. 
Thus, $d(J_s^n) = l + r^2 \bs /2$.  Matching dimensions in \eIViii\ one has
$s = (r^2 -n )\bs /2 + n + l$.  For $\bs$ irrational, one needs $n=r^2$,
and $s= r^2 + l$.  For the 
energy-momentum tensor at $s=2$, the only solution
is $n=l=1$.  Thus, the first equation in \eIxi\ is exact.  
For $s=4, 6, ...$, higher order corrections in $g$ are not ruled out
by scaling.  However, since these higher order corrections
are an issue in the massless theory, which can be folded onto
the boundary theory which is known to be integrable, we can
conclude that these possible higher order corrections do not spoil
the conservation of $Q_s$.  

The possible dimensions of the fields $R_s^n$ take the form
$d(e^{i r' \beta \phi_L \pm i \beta \phi_R} \CO'_s )$, 
where  $r'$ is an integer with $|r'| \leq n+1$,  
and the local operator $\CO'_s$ has dimension $l'$. 
Again matching dimensions in \eIViii\ , one has 
$s = (r'^2 - n -1 )\bs /2 + n + l' + 2$.  
For $\bs$ irrational, this requires $n+1 = r'^2$ 
and $s= r'^2 + l' +1$.  For $s=2$, the only solution is $n=0$, and
thus $R_2^{(+)} - R_2^{(-)} = 0$.  For $s=4$, there is also
no solution apart from $n=0$.  For $s>4$, higher order corrections
are possible.  

A similar analysis for the right moving charges $\bar{Q}_s$ 
indicates that they are all unperturbed by the defect, i.e. 
$\bar{J}_s^{(+)} (0) - \bar{J}_s^{(-)} (0) 
 = \bar{R}_s^{(+)} (0) - \bar{R}_s^{(-)} (0) = 0$. 

To summarize, in taking $\bs $ irrational, and accepting the 
folding argument by assuming that higher order corrections 
in $g$ do not spoil the conservation condition \eIxi\ 
of the massless theory, scaling arguments ensure that
the bulk perturbation for $\lambda \neq 0$ does not spoil the
conservation of $Q_s$ for at least $s=2,4$, whereas all
$\bar{Q}_s$ remain conserved.  For the energy-momentum tensor,
one has the exact equations
\eqn\eem{\eqalign{
T^{(+)} (0) - T^{(-)} (0) &= 4 \pi i g \d_t \cos \beta \phi_L ,\cr
\bar{T} ^{(+)} (0) - \bar{T}^{(-)} (0) &= 0, \cr }}
while $R_2$ and $\bar R_2$ terms are continuous across the defect.
Thus both energy and momentum are conserved.  Though
the above  scaling analysis has some gaps (we have assumed no
$\lambda$-terms in (5.7)) and doesn't
prove the integrability of the defect theory 
when $\lambda \neq 0$,  in the sequel we will
 assume that the theory \eIIi\ 
is integrable for any $\lambda$ and $g$ and is purely 
transmitting, though we emphasize we have not proven
this conclusively. 

At rational $\bs$ the above picture can be modified. 
In particular at $\bs = 1$, the free fermion point, 
$R^{(+)}_2 (0) - R^{(-)}_2 (0) $ can have non-zero
contributions at $n=1$ with $r'=l'=0$.   Thus at the free
fermion point left-moving momentum is not conserved,
and this allows for reflection in addition to transmission;
this is precisely what is found in section 6.  This is consistent
with the remarks made in the introduction, since here the bulk 
S-matrix is $-1$, and this allows {\it both} reflection and 
transmission.

\subsec{S-matrices}

We now describe 
the solution of the DYB equation in the case of
chiral sine-Gordon.
The S-matrix for solitons of 
$U(1)$ charge $\pm 1$ is as follows:
\eqn\eIIIii{\eqalign{
S_{++}^{++} (\t ) = S_{--}^{--} (\t ) &= a (\t ) ,\cr 
S_{+-}^{+-} (\t ) = S_{-+}^{-+} (\t ) &= b (\t ) ,\cr
S_{+-}^{-+} (\t ) = S_{-+}^{+-} (\t ) &= c (\t ) ,\cr
}}
where 
\eqn\eIIIiii{
a(\t ) = \sin ( \gamma (\pi + i \t )) \rho (\t ), ~~~~~ 
b(\t ) = - \sin (i \gamma \t ) \rho (\t ) , ~~~~~ 
c(\t ) = \sin (\gamma \pi ) \rho (\t ) ,} 
\eqn\eIIIiv{
\gamma = \frac{8\pi}{\beta^2} - 1 , }
and $\rho$ can be found in \ref\rzz{A. B. Zamolodchikov 
and Al. B. Zamolodchikov,
Ann. Phys. 120 (1979) 253.}.
We will consider the massless and massive cases separately.

\vskip 7pt

\noindent {\it 5.2.1 Massless Case}

\vskip 7pt

Consider first the massless case $\lambda = 0$.  The $L-R$ 
scattering, as defined by \eIIxii\ is given by 
\eqn\eIIIv{
B_{ab}^{dc} = \delta_a^d \delta_b^c ~ q^{-ab/2} , }
where 
\eqn\eIIIvi{
q = - e^{-i\gamma \pi} . }
As described in the last section, we first map the theory onto the
massless boundary sine-Gordon theory.  The reflection
S-matrices $R_a^b$ are the massless limit of the 
Ghoshal-Zamolodchikov solution \rgosh, as described in 
\ref\rhubnick{P. Fendley, 
H. Saleur, N. P. Warner, Nucl. Phys. B (1994) 577.}, and 
as such, they satisfy \eIIx\ with 
$B_{ab}^{dc} $ given by \eIIIv. 

In order to describe the scattering theory of the defect theory, 
one must answer the question:  what defect degrees of freedom 
are required in order that there is a one-to-one correspondence 
between solutions of the massless BYB \eIIx\ and the DYB \eIIv{a}?
A similar question was posed in \ref\rsal2{P. Fendley, F. Lesage
and H. Saleur, {\it A Unified Framework for the Kondo Problem
and for an Impurity in a Luttinger Liquid}, cond-mat/9510055. }.  
We introduce a defect degree of freedom $\al \in \al_0 + Z$, 
where $\al_0$ is arbitrary.  We require the defect
transmission matrix $T_{\al a}^{\beta b} $ to satisfy $U(1)$ 
charge conservation $\al + a = \beta + b$, so that $\al$ can be
interpreted as a $U(1)$ charge of the defect.  One can show that 
either of the two choices
\eqn\eIIIvii{\eqalign{
T_{\al a}^{\beta b} (\t )  & =  q^{\al a/2} ~ R_a^b (\t ), 
~~~~~a + \al = b + \beta, \cr
T_{\al a}^{\beta b} (\t )  & =  q^{-\al a/2} ~ R_a^{-b} (\t ),
~~~~~a + \al = b + \beta, \cr}}
where $R_a^b$ are the reflection S-matrices of the massless boundary
sine-Gordon theory automatically is a solution of the DYB equation if
$R$ satisfies \eIIx.  We will propose the second relation of \eIIIvii\ 
as the solution to the massless defect sine-Gordon theory.  We do so because
we expect U(1) charge to be violated in the folding process, i.e. for a
free boundary theory ($g=0$) we have
\eqn\eIIIviia{
R^+_+ = R^-_- = 0 ,~~~~~~~~ R^-_+ = R^+_- = 1 ,}
(there is maximal U(1) violation), while for a free defect theory,
we expect solitons to pass through the defect charge unchanged.

\vskip 7pt

\noindent {\it 5.2.2 Massive Case} 

\vskip 7pt

Now we take the bulk to be massive ($\lambda \neq 0$). 
As explained in the last section, the massive theory cannot be
folded into a boundary theory, so one must solve directly the 
massive DYB equation, along with the crossing, unitarity and
hermiticity constraints.  In doing this, we make the assumption
that the defect degrees of freedom in the massless and massive
cases are the same, and furthermore that the dependence on the 
defect degrees of freedom is also the same.  Thus we look
for a solution of the form 
\eqn\eIIIviii{
T_{\al a}^{\beta b} (\t ) = q^{a\al/2} T_a^b (\t ),  ~~~~~
a+ \al = b + \beta, }
where $T_a^b$ is not necessarily the same as the massless $R_a^b$. 
For scattering from the left it will be important to take 
a different ansatz where the phase depends on outgoing 
quantum numbers:
\eqn\eIIIix{
\tt_{a\al}^{b\beta} (\t ) = q^{-b\beta/2} \tt_a^b (\t ). }
There are four independent functions for each of $T$ and $\tt$:
\eqn\eIIIx{\eqalign{
T_{\al \pm}^{\al \pm} &= q^{\pm \al/2} ~ P_\pm , ~~~~~
T_{\al, \pm}^{\al \pm 2, \mp} = q^{\pm \al/2}  Q_\pm  ,\cr 
\tt_{\pm \al }^{ \pm \al } &= q^{\mp \al/2} ~ \tilde{P}_\pm , ~~~~~
T_{\pm , \al\mp 2}^{\mp, \al } = q^{\pm \al/2}  \tilde{Q}_\pm  
.\cr }}

The DYB equation leads to the equations 
\eqn\eIIIxi{\eqalign{
a (\t_{12}) P_\pm (\t_2 ) Q_\pm (\t_1 ) 
&= q~ b (\t_{12} ) Q_\pm (\t_1 ) P_\pm (\t_2 ) + c (\t_{12}) P_\pm (\t_1 )
Q_\pm (\t_2 ) \cr
a (\t_{12}) Q_\pm (\t_2 ) P_\pm (\t_1 ) 
&= q^{-1}  b (\t_{12} ) P_\pm (\t_1 ) Q_\pm (\t_2 ) + c (\t_{12}) Q_\pm (\t_1 )
P_\pm (\t_2 ) \cr
a (\t_{12}) P_\pm (\t_1 ) Q_\mp (\t_2 ) 
&= q^{-1}  b (\t_{12} ) Q_\mp (\t_2) P_\pm (\t_1 ) + c (\t_{12}) P_\pm (\t_2 )
Q_\mp (\t_1 ) \cr
a (\t_{12}) P_\pm (\t_2 ) Q_\mp (\t_1 ) 
&= q~ b (\t_{12} ) P_\pm (\t_2 ) Q_\mp (\t_1 ) + c (\t_{12}) P_\pm (\t_1 )
Q_\mp (\t_2 ) . \cr
}}
The equations satisfied by $\tilde{P}, \tilde{Q}$ are the same as above with
$q\to q^{-1}$.  

Dividing the equations \eIIIxi\ by their left hand sides, and setting
$\t_2 = 0$, one finds that one must satisfy:
\eqn\eIIIxiii{
(a - qb)(a-q^{-1} b) = c^2 . }
This requires that $q=-\exp( i\pi \gamma)$ or $-\exp(-i\pi \gamma)$. 
The two distinct solutions for these distinct values of $q$ 
correspond to taking left verses right perturbation in \eIIIi. 
To match the massless limit of the folded theory, we take 
$q = -\exp (-i\pi\gamma )$.
The general solution to the DYB equation is then 
\eqn\eIIIxiv{\eqalign{
P_\pm (\t ) &= \mu_\pm e^{-\gamma \t/2} f(\t ) , ~~~~~
Q_\pm (\t ) = \nu_\pm e^{\gamma \t /2} f(\t ) \cr
\tilde{P}_\pm (\t ) &= \tilde{\mu}_\pm e^{\gamma \t/2} \tilde{f}(\t ) , ~~~~~
\tilde{Q}_\pm (\t ) = \tilde{\nu}_\pm e^{-\gamma \t /2} \tilde{f}(\t ) ,\cr
}}
where $\mu_\pm, \nu_\pm , \tilde{\mu}_\pm , \tilde{\nu}_\pm$ are
constants, and $f, \tilde{f}$ are arbitrary. 

The hermiticity condition \eIIvd\ reads 
\eqn\eIIIxv{
\tilde{P}_\pm (\t ) = P^*_\pm (-\t ) , ~~~~~
\tilde{Q}_\pm (\t ) = Q^*_\mp (-\t ) . }
The unitarity condition \eIIiii\ then reads 
\eqn\eIIIxvi{\eqalign{
P_\pm (\t ) P^*_\pm (\t ) + Q_\pm (\t ) Q^*_\pm (\t ) &= 1 \cr
P_\pm (\t ) Q^*_\mp (\t ) + Q_\pm (\t ) P^*_\mp (\t ) &= 0 .\cr
}}
Finally, the crossing relation \eIIiv\ implies
\eqn\eIIIxvii{
P^*_\pm (\t ) = P_\mp (i\pi + \t ) , ~~~~~~~
q~ Q^*_\mp (\t ) = Q_\pm (i\pi + \t ) . }

The minimal solution to \eIIIxvi, \eIIIxvii\ of the form \eIIIxiv\ 
is the following:
\eqn\eIIIxviii{\eqalign{
P_\pm (\t ) &= e^{\pm i \xi} \frac{e^{-(\gamma \t + \nu)/2}}
{2 \cosh ( (\gamma \t + \nu - i\pi/2)/2)} ~ e^{i\delta (\t )} ,\cr
Q_\pm (\t ) &= e^{ i \tau_\pm } \frac{e^{(\gamma \t + \nu)/2}}
{2 \cosh ( (\gamma \t + \nu - i\pi/2)/2)} ~ e^{i\delta (\t )} , \cr
}}
where 
\eqn\eIIIxix{\eqalign{
e^{i\delta(\t )} &= e^{i\pi\gamma/4} \prod_{l=0}^\infty 
\frac{R_l ( \gamma \t + \nu) }
{R_l (-( \nu + \gamma \t ) ) }
\cr
R_l (x) &= 
\frac{\Gamma\(\frac{3}{4} + l\gamma - \frac{ix}{2\pi}\)
\Gamma\( \inv{4} + (l+1)\gamma - \frac{ix}{2\pi} \) }
{\Gamma\(\inv{4} + (l + \inv{2})\gamma - \frac{ix}{2\pi} \)
\Gamma\( \frac{3}{4} + (l + \inv{2}) -  \frac{ix}{2\pi} \)} 
\cr
}}
One can add to $\delta (\theta)$
a real function $r (\theta )$ such that
\eqn\eIIIxx{
r (\theta ) + r(\theta + i\pi ) = 2\pi n ,}
and still obtain a solution.  Here is the CDD ambiguity 
showing itself.
The constants $\xi, \tau_\pm$ and $\nu$ are parameters of the solution,
with $\tau_+ + \tau_- = n\pi$ where $n$ is an odd integer. 
By a gauge transformation, we can set $\tau_+ = \tau_- = n\pi/2$
for any particular odd n.  
The most significant parameter is $\nu$ which is some unknown function
of the dimensionless variables $g^2/\lambda , \beta^2 $.  

The solution 
is minimal in the sense that $\delta (\theta )$ was chosen to bring the
solution to within a phase of the massless limit 
of the folded theory and no more.  It is not surprising that we can
do so.  Once the DYB equations have constrained the phase to be
$-\exp (\pm i\pi\gamma )$, there is little freedom left for a different 
solution to arise.  Though the crossing relationship for a massive
boundary field theory,
\eqn\eIIIxixa{
R^b_{\bar a} (i\pi - \t ) = 
S^{ab}_{a'b'}(2\t - i\pi ) R^{a'}_{{\bar b}'} (\t ),}
differs significantly from its defect counterpart, its massless limit,
\eqn\eIIIxixb{
R^b_{\bar a} (i\pi - \t ) = 
q^{-ab/2} R^{a}_{\bar b} (\t ),}
differs only by a phase.  Similarly, unitarity for the boundary theory
has the same form as unitarity for the defect theory.

\subsec{Implementing the Bootstrap}

Here we will use the bootstrap to derive the transmission matrices for
the breathers.  The breathers in sine-Gordon exist as bound states of
solitons.  We can express this relation in terms of the Faddeev-Zamolodchikov 
operators.  Let $A_s (\theta )$ be the solitons ($s_1,s_2 = +/-$)
and $A_b (\theta )$ be the breathers where $b = 1, \ldots , n$ where
n is the maximum integer less than $\gamma$.  Then we have
\eqn\eIIIxxi{
f^b_{s_1 s_2} A_b (\theta ) = A_{s_1} 
(\theta + i \bar{u}^{\bar{s}_2}_{s_1{\bar b}})
A_{s_2} (\theta - i \bar{u}^{\bar{s}_1}_{s_2{\bar b}}) ,}
where $\bar{s} = -s$ is the charge conjugate of $s$.  The $\bar{u}$'s are
given by 
\eqn\eIIIxxii{
\bar{u}^{s'}_{sb} = \pi - u^{s'}_{sb} , }
where $i u^{s'}_{sb}$ 
is the location of the soliton pole in the soliton-breather
scattering matrix.  Here we have
\eqn\eIIIxxiii{
u^{\pm}_{\pm b} = {\pi \over 2} + {n\pi\over 2\lambda} .}
The $f$'s are related to the residues of the poles in the soliton-anti-soliton
scattering matrices.  If $u^{b}_{+-}$ is the location of the pole in 
$S^{\pm\mp}_{+-}$, indicative of breather b, the $f^{b}_{\mp\pm}$ are
defined via
\eqn\eIIIxxiv{
S^{\pm\mp}_{+-} (\theta ) \sim i {f^b_{+-} f^{\pm\mp}_b \over \theta
- iu^b_{+-}},}
where there is no summation on $b$.  $f^b_{s_1s_2}$ thus represents
the probability amplitude that the bound state forms from a soliton
pair.  It is then easy to show that
\eqn\eIIIxxv{\eqalign{
f^b_{+-} f^{+-}_b & = (-1)^b S_o ;\cr
f^b_{-+} f^{+-}_b & = S_o ;\cr
f^b_{\pm\pm} & = 0,}}
where $S_o$ is some constant.  Hence $f^b_{+-}/f^b_{-+} = (-1)^b$.
This last relation is all we will need for the computation of the breather 
transmission matrices.

The transmission bootstrap equation is
\eqn\eIIIxxvi{
f^b_{s_1s_2} T^{\gamma d}_{\alpha b} (\theta )
 = f^d_{g_3g_2} 
T^{\beta g_2}_{\alpha s_2} (\theta - i\bar{u}^{\bar{s}_1}_{s_2\bar{b}})
T^{\gamma g_3}_{\beta s_1} (\theta + i\bar{u}^{\bar{s}_2}_{s_1\bar{b}}).}
This is derived through the consistency of \eIIIxxi\ scattering through
the defect.  In fact it is more general than indicated: 
$b$ need not be a breather and 
$s_1,s_2$ need not be solitons for the equation to hold.  

Now taking 
$b$ to be some breather and $s_1=+,s_2=-$, we can write an explicit expression
for the breather transmission matrix:
\eqn\eIIIxxvii{
T^{\alpha b}_{\alpha b} (\theta ) = P_-(\theta -i\bar{u}^-_{-b})
P_+(\theta + i\bar{u}^+_{+b}) + (-1)^{b}q^{-1}Q_-(\theta -i\bar{u}^-_{-b})
Q_+(\theta + i\bar{u}^+_{+b}) .}
As expected, the breather scattering matrix does not depend on the
parameter $\alpha$.  With this, we can explicitly check if 
$T^{\alpha b}_{\alpha b} (\theta) $ satisfies crossing and unitarity.
We find that both
\eqn\eIIIxxviii{\eqalign{
& {T^*}^{\alpha b}_{\alpha b} (\theta) = 
T^{\alpha b}_{\alpha b} (i\pi + \theta);\cr
& T^{\alpha b}_{\alpha b} (\theta) {T^*}^{\alpha b}_{\alpha b} (\theta) =1,}}
hold without putting any further constraints on $\delta (\theta )$ in
\eIIIxviii\ and \eIIIxix\ .  

\newsec{The Free Fermion Point}

In this section we will consider 
the chiral defect SG theory at the free fermion point
$\beta^2 = 4\pi$, where we can derive the defect scattering
matrices from an action.  For the reasons indicated at the end of
section 5.1, we expect that this theory will exhibit reflection
in addition to transmission in the massive case $\lambda \neq 0$, 
and this is what we find.  In the massless limit we recover the
purely transmitting S-matrix of section 5.2.  This suggests the
massive transmission S-matrix of section 5 only has
significance in its massless limit, since it was obtained 
assuming only transmission.
However we have not absolutely 
ruled out that
an alternative massive fermion action may still be purely
transmitting with massive transmission given by 
$T^{\beta b}_{\alpha a} (\t , \beta^2 = 4\pi)$ of the last section.

In terms of fermions, the action \eIIi\ can be expressed as 
\eqn\eVIi{\eqalign{S &= S_{{\rm bulk}} + S_{{\rm free}} + S_{{\rm defect}} \cr
& = {1 \over 8\pi} \sum_{i = \pm} \int dx dt ~
\psi^{(i)}_+ \partial_{\bar z} \psi^{(i)}_- + 
\psi^{(i)}_- \partial_{\bar z} \psi^{(i)}_+ +
\bar{\psi}^{(i)}_+ \partial_{z} \bar{\psi}^{(i)}_- + 
\bar{\psi}^{(i)}_- \partial_{z} \bar{\psi}^{(i)}_+ \cr
& \hskip 50pt + 2im 
(\bar{\psi}^{(i)}_+\psi^{(i)}_- - \psi^{(i)}_+\bar{\psi}^{(i)}_-) \cr
& -{i \over 8\pi} \int dt ~ \psi^{(+)}_-\psi^{(-)}_+ +
\bar{\psi}^{(-)}_+\bar{\psi}^{(+)}_- + \psi^{(+)}_+\psi^{(-)}_- +
\bar{\psi}^{(-)}_-\bar{\psi}^{(+)}_+ \cr
& - {g \over 4\pi} \int dt ~\psi^{(-)}_+ a_- + \psi^{(+)}_+ a_+
+ a_+ \psi^{(-)}_- + a_-\psi^{(+)}_- 
- {1 \over 2\pi} \int dt ~ a_+ \partial_t a_- .}}
Here we have used the standard map between fermions and bosons at
$\beta^2 = 4 \pi $: $\psi_{\pm} = \exp (i\phi_L)$ and 
$\bar{\psi}_{\pm} = \exp (\mp\phi_R$) 
\ref\cole{S. Coleman, Phys. Rev. D 11 (1975) 2088.} 
\ref\mandel{S. Mandelstam, Phys. Rev. D 11 3026 (1975).} .
The bulk term represents, locally,
the massive sine-Gordon action.  However, we have divided it into two parts:
one for the theory in the region $(x>0)$ (terms with the superscript $(+)$)
and one for the region $(x<0)$ (terms with the superscript $(-)$).  Because
the two regions $(+)$ and $(-)$ are decoupled, we need to add to the
action a term which insures continuity, i.e.
\eqn\eVIii{
\psi_\pm^{(-)} - \psi^{(+)}_\pm = \bar{\psi}^{(-)}_\pm - 
\bar{\psi}^{(+)}_\pm = 0,}
in the absence of a defect ($g=0$).  This term is $S_{{\rm free}}$.  The
defect term is constructed in analogy to what was done
 in \ref\amer{M. Ameduri,
R. Konik, and A. LeClair, Phys. Lett. B 354 (1995) 376.}.  There Fermi fields
were coupled to fermionic
defect modes, $a_\pm$.  Although the origin of the $a_\pm$ is more 
obscure here, we must have the same thing, if only to insure the action is
bosonic.  The specific choice of mixing and matching the $a_\pm$'s and
the $\psi_\pm$'s is determined by unitarity, by the need to have terms
which both violate and preserve U(1) charge, and in part by trial and
error (so as to obtain, in the massless limit, the solution of the DYB
equations in section 5).
The defect degrees of freedom introduced in section 5 can be understood
as arising from the zero modes $a_\pm$.  
To derive the scattering matrices for this action, we begin by
writing down the equations of motion.  Varying with respect to
$\psi_\pm^{(+),(-)}$, $\bar{\psi}_\pm^{(+),(-)}$, and $a_\pm$, and
then eliminating the $a_\pm$ modes, leaves us with three equations,
all evaluated at $x=0$:
\eqn\eVIiii{\eqalign{
0 & = i \partial_t (\psi^{(+)}_+ - \psi^{(-)}_+) + g^2 (\psi^{(+)}_+ - 
\psi^{(-)}_-) ,\cr
0 & = \psi^{(+)}_- - \psi^{(-)}_- + \psi^{(+)}_+ - \psi^{(-)}_+ ,\cr
0 & = \bar{\psi}^{(+)}_\pm - \bar{\psi}^{(-)}_\pm .}}
The first two equations describe interactions among the chiral fermions.
The last equation is nothing more than the continuity of the anti-chiral
fermions.

To determine the scattering matrices, we employ the following mode
expansions for the fermions:
\eqn\eVIiv{\eqalign{
\psi_+ & = \sqrt{m} \int^\infty_{-\infty} {d\theta \over 2 \pi i}
e^{-\t /2} \( A_-(\t ) e^{-m(ze^{-\t } +\bar z e^{\t })} -
A_+^\dagger (\t ) e^{m(ze^{-\t } +\bar z e^{\t })} \) \cr
\bar{\psi}_+ & = -i \sqrt{m} \int^\infty_{-\infty} {d\theta \over 2 \pi i}
e^{\t /2} \( A_-(\t ) e^{-m(ze^{-\t } +\bar z e^{\t })} +
A_+^\dagger (\t ) e^{m(ze^{-\t } +\bar z e^{\t })} \) ,}}
where $\psi_- = \psi^\dagger_+$ and $\bar{\psi}_- = \bar{\psi}^\dagger_+$.
The A's satisfy the following anti-commutation relations:
\eqn\eVIv{
\{ A_\pm (\t ), A^\dagger_\mp (\t ') \} = 4\pi^2 \delta (\t - \t ').}
We now substitute the expansions into the equations of motion, Fourier
decompose, and express the equations in the form:
\def\d{{\bf D}}
\def\ap{A^{\dagger (+)}}
\def\am{A^{\dagger (-)}}
\eqn\eVIvi{\eqalign{
\d \ap_a (\t ) & = T^b_a (\t ) \am_b (\t ) \d + R^b_a (\t ) \d \ap_b (-\t );\cr
\am_a (\t )\d  & = {\tilde T}^b_a (-\t ) \d \ap_b (-\t ) + 
{\tilde R}^b_a (-\t ) \am_b (-\t )\d ,}}
or upon expanding out
\eqn\eVIvi{\eqalign{
\d \ap_\pm (\t ) & = P_\pm (\t ) \am_\pm (\t ) \d + 
Q_\pm (\t ) \am_\mp (\t ) \d + \cr
& \hskip 1cm M_\pm (\t ) \d \ap_\pm (-\t ) + N_\pm (\t ) \d \ap_\mp (-\t );\cr
\am_\pm (\t ) \d & = {\tilde P}_\pm (-\t ) \d \ap_\pm (\t ) + 
{\tilde Q}_\pm (-\t ) \d \ap_\mp (\t ) + \cr
& \hskip 1cm {\tilde M}_\pm (-\t ) \am_\pm (-\t )\d  + 
{\tilde N}_\pm (-\t ) \am_\mp (-\t )\d .}}
These equations define the S-matrices for the theory.  $P_\pm$ and
$Q_\pm$ represent transmission across the defect while $M_\pm$ and
$N_\pm$ represent reflection from the defect.  We find the matrices to be
\eqn\eVIvii{\eqalign{
P_\pm (\t ) &= {\tilde P}^*_\pm (-\t ) = 
{f(\t ) + e^{2\t } - 1 \over f(\t )};\cr
Q_\pm (\t ) &= {\tilde Q}^*_\pm (-\t ) = 
{1 - e^{2\t } \over f(\t )};\cr
M_\pm (\t ) &= {\tilde M}^*_\pm (-\t ) = 
-{2\sinh (\t ) \over f(\t )};\cr
N_\pm (\t ) &= {\tilde N}^*_\pm (-\t ) = 
{2\sinh (\t ) \over f(\t )}, }}
where 
\eqn\eVIviii{
f(\t ) = -{4im \over g^2} \cosh (\t )\sinh^2(\t ) + 2\sinh (2\t ).}
The matrices do not differentiate between U(1) charge because of our
choice of $S_{\rm free}$.  If we wrote $S_{\rm free}$ as
\eqn\eVIix{
S_{\rm free} = -{i \over 8\pi} \int dt ~ 
e^{i\alpha} \psi^{(+)}_-\psi^{(-)}_+ +
e^{i\beta} \bar{\psi}^{(-)}_-\bar{\psi}^{(+)}_- + 
e^{-i\alpha} \psi^{(+)}_+\psi^{(-)}_- +
e^{-i\beta} \bar{\psi}^{(-)}_-\bar{\psi}^{(+)}_+ ,}
the $\pm$ matrices would be distinguished by a phase.

Because the theory is both reflecting and transmitting, crossing and
unitarity are altered.  Unitarity becomes
\eqn\eVIx{\eqalign{
1 & = T^b_a(\t ){\tilde T}^c_b (-\t ) + R^b_a (\t ) R^c_b (-\t ) ; \cr
0 & = T^b_a (\t ) {\tilde R}^c_b (-\t ) + R^b_a(\t ) T^c_b (-\t ),}}
in addition to two more equations where ${\tilde T}(-\t )$ and 
${\tilde R}(-\t )$ are interchanged 
with $T(\t )$ and $R(\t )$.  Crossing keeps its
transmission component in (4.4) and adds a reflection piece:
\eqn\eVIxi{\eqalign{
R^b_{\bar a} (\t ) = - R^{a}_{\bar b} (i\pi - \t );\cr
{\tilde R}^b_{\bar a} (\t ) = - {\tilde R}^{a}_{\bar b} (i\pi - \t ). }}
One can easily check that the scattering matrices derived above satisfy 
these relations.  

As promised, the massless limit of these matrices match the
scattering matrices in 5.34.  To find the massless limit of the 
matrices for scattering of the left movers from right to left 
take $\t \rightarrow \t_L - \alpha$ and
let $\alpha \rightarrow \infty$, $m \rightarrow 0$ while holding
$\mu = me^{\alpha}$ constant.  We so find
\def\tln{\theta_L + \nu}
\def\tl{\theta_L}
\eqn\eVIxii{\eqalign{
P_\pm (\tl ) &= 
{e^{i\pi/4} e^{-(\tln )/2} \over 2\cosh ((\tln - i\pi/2)/2)},\cr
Q_\pm (\tl ) & 
= {e^{-i\pi/4} e^{(\tln )/2} \over 2\cosh ((\tln - i\pi/2)/2)},\cr
M_\pm (\tl ) & = N_\pm(\tl ) = 0 .}}
Here we have identified $\nu$ with $-\log (\mu/2g^2 )$.  To obtain
an exact match with section 5, we need to fix our gauge by setting
$\tau_+ = \tau_- = -\pi/2$.

\newsec{Ising Model}

Having considered a chiral defect theory for a Dirac fermion, it is
natural to also treat a Majoranna fermion, i.e. the fermionic representation
of the Ising model.  As with the Dirac fermion, we find a massive Majoranna
fermion has both transmission and reflection S-matrices.  And
again, we find that in the massless limit, the reflection matrices vanish.

Defect lines in Ising models have been considered before
(\ref\sov{R. Bariev, Sov. Phys. JETP 50 (1979) 613.}
\ref\mcc1{B. McCoy and J. H. H. Perk, Phys. Rev. Lett. 44 (1980) 840.}
\ref\mcc2{B. McCoy and J. H. H. Perk, The Riemann Problem, Complete
Integrability and Arithmetic Applications 925 (1982) 12.}), but in
a different form.  Previously, defects as  perturbations in the energy
were studied.  In terms of fermions, a energy perturbation is represented
by $\psi\bar\psi$. Here instead we consider defects as perturbations
in a single $\psi$.  Whereas energy perturbations have ready interpretations
in terms of spin operators in a lattice formulation, a chiral perturbation
does not.  Its physical interpretation comes only in folding the theory onto
boundary Ising.

The action for the Ising model can be expressed in the same form
as \eVIi\ ,
\eqn\eVIIi{\eqalign{S &= S_{{\rm bulk}} + S_{{\rm free}} + S_{{\rm defect}} \cr
= & {1 \over 8\pi} \sum_{i = \pm} \int dx dt ~
\( \psi^{(i)} \partial_{\bar z} \psi^{(i)} + 
\bar{\psi}^{(i)} \partial_{z} \bar{\psi}^{(i)} + 
2im \psi^{(i)}\bar{\psi}^{(i)} \) \cr
& - {i \over 8\pi} \int dt ~ \( \psi^{(+)}\psi^{(-)} +
\bar{\psi}^{(+)}\bar{\psi}^{(-)} \) \cr
& + {i g \over 4\pi} \int dt ~ \( a\psi^{(+)} - \psi^{(-)} a \)
- {1 \over 4\pi} \int dt ~ a \partial_t a .}}
With the Dirac fermion, the defect modes, $a_\pm$, were indexed to reflect
the U(1) charge carried by the fermionic fields.  Here there fields have
no charge and so the defect mode, $a$, goes unstructured.

As before, we determine the scattering matrices by varying the action
and writing down the equations of motion at $x=0$:
\eqn\eVIIii{\eqalign{
0 & = \partial_t (\psi^{(+)} - \psi^{(-)}) -ig^2 (\psi^{(+)} + \psi^{(-)});\cr
0 & = \bar{\psi}^{(+)} - \bar{\psi}^{(-)}.\cr}}
The first equation describes the interaction of the chiral fermion,
while the second equation simply enforces the continuity of the
anti-chiral fermion across the defect line.  We employ the 
following mode expansions:
\eqn\eVIIiii{\eqalign{
\psi & = \sqrt{m} \int^\infty_{-\infty} {d\theta \over 2 \pi i}
e^{-\t /2} \( A(\t ) e^{-m(ze^{-\t } +\bar z e^{\t })} -
A^\dagger (\t ) e^{m(ze^{-\t } +\bar z e^{\t })} \) \cr
\bar{\psi} & = -i \sqrt{m} \int^\infty_{-\infty} {d\theta \over 2 \pi i}
e^{\t /2} \( A(\t ) e^{-m(ze^{-\t } +\bar z e^{\t })} +
A^\dagger (\t ) e^{m(ze^{-\t } +\bar z e^{\t })} \) ,}}
Here $\{ A(\t ),A^\dagger (\t ')\} = 4\pi^2 \delta (\t -\t ')$.
Substituting these expansions into the equations of motion, 
and reducing them to the form
\eqn\eVIIiv{\eqalign{
\d \ap (\t ) & = T (\t )\am \d +  R (\t ) \d \ap (-\t ) \cr
\am (\t ) \d & = {\tilde T} (-\t ) \d \ap (\t ) + 
{\tilde R} (-\t ) \am (-\t )\d ,\cr}}
leads to the result
\eqn\eVIIv{\eqalign{
T (\t ) & = {\tilde T}^*(-\t ) = {m\cosh (\t ) + ig^2 \over m\cosh (\t )
+ ig^2 \coth (\t )} ;\cr
R(\t ) & = {\tilde R}^*(-\t ) = -{ig^2 \over m\cosh (\t )\sinh (\t )
+ ig^2 \cosh (\t )} .}}
In the massless limit, $R(\t )$ and $T(\t )$ become
\eqn\eVIIva{\eqalign{
T(\theta_L ) & = {\mu e^{-\theta_L} + i 2 g^2 \over 
\mu e^{-\theta_L} - i2g^2},\cr
R(\theta_L ) & = 0.}}
As expected the reflection amplitude, $R(\theta_L)$, vanishes.  The massless
transmission matrix can be compared with the massless limit of the reflection
matrix, $R_B$, derived in \rgosh~for boundary Ising.  In this case we find
\eqn\eVIIvb{
R_B(\theta_L ) 
= -i {\mu e^{-\theta_L} + i h^2 \over \mu e^{-\theta_L} - i h^2},}
where $h$ is the boundary magnetic field.  By identifying $2g^2$ with
$h^2$, a phase is the 
sole difference between the two massless limits.
This phase arises from the difference between the boundary and defect crossing
relations.

In this case unitarity takes the form,
\eqn\eVIIvi{\eqalign{
1 & = T(\t ) {\tilde T}(-\t ) + R(\t )R(-\t ) ;\cr
0 & = T(\t ) {\tilde R}(-\t ) + R(\t )T(-\t ).}}
while crossing demands,
\eqn\eVIIvii{
T(\t ) = {\tilde T}(i\pi - \t ), ~~~~
R(\t ) = -R (i\pi - \t ), ~~~~
{\tilde R}(\t ) = - {\tilde R} (i\pi - \t ) .}
It easily seen that the scattering amplitudes satisfy these equations.

\newsec{Conclusions}

We have outlined a general approach to integrable chiral defect
theories, and have applied this to a chiral defect perturbation
of the sine-Gordon theory.  

A nice feature of purely transmitting theories of defects
is that they allow one to consider multiple defects rather easily,
and consequently a finite density of impurities. 
We hope to turn to this in the future.

\bigskip
\centerline{\bf Acknowledgments} 

We would like to thank F. Lesage, A. Ludwig,
 and H. Saleur for useful discussions. 
This work is supported both by the National Science Foundation,
in part through the National Young Investigator Program, 
and under Grant No. PHY94-07194, and the National Science and
Engineering Research Council of Canada, through an NSERC PGS B fellowship.

\listrefs
\bye